\documentclass[aps,twocolumn,superscriptaddress,nofootinbib,preprintnumbers]{revtex4}
\usepackage[dvips]{color}
\usepackage[normalem]{ulem}
\usepackage{amsmath}
\usepackage{enumerate}
\usepackage{amsfonts}
\usepackage{epsfig}
\usepackage{yfonts}

\usepackage{graphicx}
\usepackage{bm}

\newcommand\comment[1]{}

\def\({\left(}
\def\){\right)}
\def\[{\left[}
\def\]{\right]}
\def\<{\langle}
\def\>{\rangle}
\newcommand{\be}{\begin{equation}}
\newcommand{\ee}{\end{equation}}
\newcommand{\bea}{\begin{eqnarray}}
\newcommand{\eea}{\end{eqnarray}}
\newcommand{\bwt}{\begin{widetext}}
\newcommand{\ewt}{\end{widetext}}

\newcommand{\bi}{\begin{itemize}}
\newcommand{\ei}{\end{itemize}}
\newcommand{\ben}{\begin{enumerate}}
\newcommand{\een}{\end{enumerate}}
\newcommand{\bca}{\begin{cases}}
\newcommand{\eca}{\end{cases}}
\newcommand{\bln}{\begin{align}}
\newcommand{\eln}{\end{align}}
\newcommand{\bst}{\begin{split}}
\newcommand{\est}{\end{split}}

\newcommand{\bI}{\mathbf{I}}
\newcommand{\bII}{\mathbf{II}}

\renewcommand{\Im}{\textrm{Im}\,}

\newcommand{\cO}{\mathcal{O}}

\def\spinor{\zeta}

\def\mspinor{m}  %_\zeta}

\begin{document}

\title{Fermi arcs from holography}

\author{David Vegh  \\
 \normalsize {\it Simons Center for Geometry and Physics, Stony Brook University, Stony Brook, NY 11794-3636}
 \\  % \date{June 16, 2010}   %(\today)
}

%
%\date{June 09, 2010}
%\hskip.2in
\begin{abstract}

In this paper, we find mechanisms for the generation of Fermi arcs using the gauge/gravity correspondence. The gravity background is taken to be a charged black hole with vector hair in asymptotically $AdS_4$ spacetime. The response function of fermion probes exhibits a p-wave gap in the dual superconductor. We couple the fermions to a charged rank-two antisymmetric field. Assuming that its spatial components condense, a novel type of open Fermi surface is produced.
We derive an analytical formula for the Green's function and study its unique properties.
The results are confirmed by separate numerical computations.

In the Appendix, we study the effect of a neutral scalar field on the fermionic spectral functions. A suitable interaction term shifts the original spin-up / spin-down Fermi momenta in opposite directions and thus the two nodal points of the p-wave gap extend into Fermi pockets.

\end{abstract}

\maketitle

\section{Introduction}

Strongly coupled systems exhibit various interesting phenomena which often cannot be understood using intuition from weakly coupled physics. Among these are many unconventional properties of high-temperature superconductors \cite{Bednorz}. Understanding these materials remains one of the major challenges in physics. %testing our ability to describe nature.

In the {\it superconducting phase} of cuprates, an anisotropic energy gap opens up near the Fermi level. The interesting physics is essentially two-dimensional and the order parameter has a d-wave symmetry. This means that the gap is approximately a $\textrm{cos}(2\theta)$ function of the $\theta$ angle in momentum space and thus it vanishes at the four nodal points on the diagonals of the Brillouin zone. The phase transition temperature ($T_c$) depends on the amount of electron or hole doping ($x$). At zero doping, these materials are antiferromagnetic Mott insulators. The superconducting phase covers a dome-shaped region in the $x-T$ phase diagram. Optimal doping is achieved when $T_c$ is maximal. Underdoping and overdoping refer to superconductors with doping levels below and above optimal doping, respectively.

Underdoped cuprates  exhibit fascinating phenomena including charge stripes, a large Nernst effect, unusual specific heat, spin susceptibility and transport properties. Above $T_c$, in the metallic {\it pseudogap phase}, the density of states near the Fermi energy is partially suppressed and the four nodal ``Fermi points'' extend into Fermi arcs: gapless excitations in disconnected arc-shaped regions of momentum space.
The length of the arcs grows with temperature and doping until a large connected Fermi surface is recovered. The existence of open Fermi arcs is rather interesting since in conventional metals Fermi surfaces are closed boundaries separating occupied and unoccupied states. The origin of Fermi arcs is not a settled question, although different models exhibiting similar features have been proposed\footnote{For angle-resolved photoemission spectroscopy (ARPES) and quantum oscillation data, see {\it e.g.} \cite{2006PhRvL..96j7006C, 2006NatPh...2..447K, 2007Natur.447..565D, 2008PhRvL.101m7002K} and the reviews \cite{2002cond.mat..6217C, 2002cond.mat..9476C, 2003RvMP...75..473D}. For theoretical models, see \cite{2004cond.mat.10445L, 2007PhRvB..76q4501N} and references therein.}.

Motivated by cuprate phenomenology, we will study anisotropic Fermi surfaces using the gauge/gravity correspondence \cite{Maldacena:1997re, Gubser:1998bc, Witten:1998qj}\footnote{For recent developments in the topic of holographic Fermi liquids, see \cite{Lee:2008xf, Liu:2009dm, Cubrovic:2009ye, Faulkner:2009wj, Denef:2009kn, Faulkner:2010tq}. For holographic superconductors, see \cite{Gubser:2008px, Hartnoll:2008vx, Hartnoll:2008kx, Horowitz:2009ij, Gubser:2009cg, Faulkner:2009am, Gubser:2009dt, Chen:2009pt, Gubser:2010dm}. For further references, see the recent reviews \cite{Hartnoll:2009sz, Herzog:2009xv, McGreevy:2009xe, Sachdev:2010ch, Faulkner:2010da, Horowitz:2010gk}.}. {\it The purpose of the paper is to find Fermi arc-like phenomena using ingredients which are natural from a holographic point of view. }

In Section II, we describe the (3+1)-dimensional gravity background.
In order to introduce anisotropy in the system, we consider an anti-de Sitter black hole with a condensed vector field. A non-zero spatial component of the vector field breaks rotational invariance in the boundary theory and gives a p-wave holographic superconductor\footnote{Alternatively, one can substitute the vector field with spin-two (or higher spin) fields as in the recent papers \cite{Chen:2010mk, Benini:2010qc}. In the latter paper, Fermi arcs were due to temperature broadening.}. Probe fermions are coupled to the vector field. The p-wave gap is seen in their response functions: there are low-energy excitations near the two nodal points.

In Section III, we couple the fermions to an antisymmetric tensor field background. The fermionic response functions exhibit Fermi arcs.
Note that these arcs exist in the superconducting phase. Using approximations whose results are confirmed by numerical computations, we derive an analytical formula for the Green's function.
We show that the antisymmetric coupling enhances temperature broadening in the spectral function. More interestingly, it decreases the distance between the quasiparticle poles on the complex frequency plane.
{\it Thus, it provides a heretofore unknown mechanism for generating Fermi surfaces that are not closed. }

In the Appendix, we describe a different model where the fermions couple to a neutral scalar field. The coupling behaves as an effective mass term and it shifts the original ``spin-up'' and ``spin-down'' Fermi momenta in different directions. As a result, near the p-wave nodal directions the two gapless points extend into ``Fermi pockets''.

\section{The setup}

\subsection{Anisotropic background}

Let us consider a massive $W_I$ vector field in the bulk. This will play the role of the p-wave order parameter\footnote{I thank Michal Heller for collaboration.} \cite{Gubser:2008wv, Roberts:2008ns, Basu:2008st, Ammon:2009xh}. We assume that it has charge $2q$ under the  $U(1)$ gauge symmetry. The action takes the form,
\bea
%  \nonumber
\label{u1action}
  S = {1 \over 2\kappa^2} \int d^4 x \sqrt{-g} \left[ R + {6 \over L^2} - {L^2 \over 4 g^2}
  F_{MN}F^{MN} \qquad
  \right.   \\
  \left.  - G_{MN}G^{MN} -m_W^2 W_N W^N \right]  \nonumber
\eea
with $G_{MN} = D_{[M} W_{N]}$ and $D_I W_J = \partial_I W_J + i (2q) A_I W_J$.

\noindent
For the condensed phase, we take the ansatz,
\be%\label{eq:metric}
  \nonumber
%  ds^2=-g(r) e^{-\chi(r)} dt^2+{dr^2\over g(r)}+r^2 f(r)^2 dx^2 + {r^2 \over f(r)^2} dy^2
%  ds^2=-g(r) e^{-\chi(r)} dt^2+{dr^2\over g(r)}+  {r^2 \over f(r)^2} dx^2 + r^2 f(r)^2 dy^2
  ds^2=-g(r) e^{-\chi(r)} dt^2+{dr^2\over g(r)}+  r^2\left( { dx^2 \over f(r)^2}  +   f(r)^2 dy^2 \right)
\ee
\be
   A=\phi(r) dt,  \quad W  = W_{y}(r) dy~~.
    \nonumber
\ee
The $f, g, \chi, \phi$ and $W_y$ functions can be determined numerically from the equations of motion \cite{michal} such that the metric describes a static asymptotically $AdS_4$ spacetime with a charged black hole of Hawking temperature $T$ and a condensed vector field hair. The horizon is located at $r=r_*$. The metric is anisotropic in the spatial directions.

\subsection{Spinor probes}

Let us now introduce two\footnote{In even dimensions, we could use a single spinor instead and couple it to its charged conjugate as in \cite{Faulkner:2009am}.} probe bulk Dirac fields $\spinor_1$ and $\spinor_2$ with opposite $U(1)$ charges, $\pm q$, under the $U(1)$ gauge symmetry.
The fermionic action takes the form\footnote{In order to write down physically interesting interaction terms, as an organizing principle, it is useful to think of (\ref{u1action}) as the result of a broken ``pseudospin'' $SU(2)$  bulk gauge symmetry. The $U(1)$ bulk gauge symmetry is interpreted as the unbroken diagonal subgroup and $W$ is the pseudospin W-boson. The $\spinor_i$ fields then form a pseudospin doublet.},
\bea
 \nonumber
  %S_\textrm{probe} =
  S_0= \int d^4 x \,\sqrt{-g}  \left[i \overline{\spinor}_a \left( \Gamma^M \overleftrightarrow
  D_M - \mspinor \right) \spinor_a  \qquad  \right. \\
   \left. +   \eta  W_{N} \overline{\spinor}_2  \Gamma^{N}  \spinor_1  + h.c.
  \right]
   \nonumber
\eea
where $a=1,2$ and $\eta$ parametrizes the coupling of the fermions to the p-wave order. We are going to use the following basis for gamma matrices,
\be
\Gamma^{\underline{r}}=
\begin{pmatrix}
-\sigma^3  &   0 \\
0 & -\sigma^3
\end{pmatrix}
\qquad \Gamma^{\underline{t}}=
\begin{pmatrix}
i \sigma^1  &   0 \\
0 & i\sigma^1
\end{pmatrix}
\nonumber
\ee
\be
\Gamma^{\underline{x}}=
\begin{pmatrix}
-\sigma^2  &   0 \\
0 & \sigma^2
\end{pmatrix}
\qquad \Gamma^{\underline{y}}=
\begin{pmatrix}
0 &  \sigma^2 \\
\sigma^2 & 0
\end{pmatrix} .
 \nonumber
\ee
 For convenience, we also give the following matrices,
\be
\nonumber
\Gamma^{\underline{5}}=
\begin{pmatrix}
0 & i\sigma^2   \\
-i\sigma^2 & 0
\end{pmatrix}
\qquad \Gamma^{\underline{x}}\Gamma^{\underline{y}}=
\begin{pmatrix}
0 &  -1 \\
1 & 0
\end{pmatrix} .
\ee

\noindent
Since in the above background the $\omega_{ab M}$ spin connection satisfies
\begin{equation} \frac{1}{4} \omega_{ab M} e^M_c \Gamma^c
 \Gamma^{ab} = \frac{1}{4} \Gamma^r \partial_r \ln \left( - g g^{rr} \right) ,
  \nonumber
\end{equation}
we can rescale the Dirac fields and remove the spin connection from the
equations. For a given $\omega, \vec k$  mode, we introduce the notation in the above gamma matrix basis,
\be
 \spinor_a =:   ( - g g^{rr} )^{-1/4} e^{-i \omega t + i \vec k \vec x}  \begin{pmatrix} \lambda_{a}(\omega, \vec k) \\ \chi_{a}(\omega, \vec k)
  \end{pmatrix}  .
\ee
The four-component spinor $\spinor$ has been split into the rescaled $\lambda$ and $\chi$ two-component spinors.

\subsection{Green's function}

If we restrict the spinor momentum to be in the $x$ direction (perpendicular to the condensed field), then the equations decouple into two sets of equations containing $\lambda_{1}, \chi_{2}$ and $\lambda_{2}, \chi_{1}$, respectively\footnote{I thank Hong Liu for pointing out that this can also be done in the general case by applying a change of basis.}. Thus, we can consider these variables separately. In the rest of the paper, we will focus on $\lambda_{1}$ and $\chi_{2}$, suppressing their 1, 2 indices. ($\lambda_{2}$, $\chi_{1}$ can be treated similarly.)
Let us combine these spinors into the four-component {\it Nambu-Gor'kov spinor} $\psi=\binom{\lambda_{1}}{\chi_{2}}$. The Dirac equation,
%The Dirac equation for the {\it Nambu-Gor'kov spinor} %$\psi:=\binom{\lambda_{1}}{\chi_{2}}$,
\be
\nonumber
\begin{pmatrix} D_r (k, q) + \sqrt{g^{tt}} \omega \sigma^1 &  \  i\eta \sqrt{g^{yy}} W_y \sigma^2    \\
i\eta \sqrt{g^{yy}} W_y \sigma^2     &   D_r(-k, -q)+\sqrt{g^{tt}} \omega \sigma^1 \end{pmatrix} \psi =0 .
% \binom{\lambda_{1}}{\chi_{2}} =0
\ee
Here we used the notation,
\be \nonumber
 D_r (k, q) \equiv -\sqrt{g^{rr}} \sigma^3 \partial_r  - \mspinor - \sqrt{g^{xx}} i \sigma^2 k +
\sqrt{g^{tt}} \sigma^1 q A_t \ .
\ee
The off-diagonal terms are subdominant at the UV and IR boundaries. The two linearly independent solutions with ingoing boundary conditions at the horizon will be denoted by ${\bI}$ and ${\bII}$.
At the UV $AdS_4$ boundary, the two independent
solutions for a two-component spinor ($\theta = \lambda  \ \textrm{or} \ \chi$) are,
\be
  \theta(r) \stackrel{r \rightarrow \infty}{\longrightarrow}
       \theta_+ r^{mR} \binom{ 0}{1}
    + \theta_- r^{-mR} \binom{ 1}{ 0} . \nonumber
\ee
The Green's function matrix is given by \cite{Henningson:1998cd, Mueck:1998iz, Iqbal:2009fd, Faulkner:2009wj, Faulkner:2009am},
\begin{equation}
\label{eq:amatrix}
\underbrace{ \begin{pmatrix} \lambda^{\bI}_- &   \lambda^{\bII}_- \\
 \chi^{\bI}_- & \chi^{\bII}_-  \end{pmatrix}  }_\textrm{expectation value $(\mathcal{E})$}
\hskip -0.4cm =  \begin{pmatrix} G_{\cO_1 \cO^\dagger_1} &   G_{\cO_1 \cO_2^\dagger} \\
G_{\cO_2 \cO_1^\dagger} & G_{\cO_2 \cO_2^\dagger} \end{pmatrix}
\underbrace{\begin{pmatrix} \lambda^{\bI}_+ &   \lambda^{\bII}_+ \\
 \chi^{\bI}_+ & \chi^{\bII}_+   \end{pmatrix}  }_\textrm{source $(\mathcal{S})$}
%  \nonumber
\end{equation}
where
\be
  G_{XY}(\omega, \vec k) = i \int d\vec x \, dt \, e^{-i \omega t + i \vec k \vec x}  \theta(t) \langle \{ X(x,t), Y(0,0) \} \rangle
  \nonumber
\ee
is the retarded correlator and $\cO_i$ denote the fermionic boundary operators dual to the bulk spinor fields.

The condensed vector field induces a mixing of positive and negative frequency modes of the probe spinors. This mixing is maximal when their momenta lie in perpendicular ($y$) direction. The eigenvalue repulsion between particles and holes at $\omega=0$ produces a p-wave gap in the fermion spectral function, see FIG. \ref{tripfig} (i).

\section{Fermi arc}

\label{sec:bfield}

In this section, we describe an antisymmetric coupling that can be used to produce a Fermi arc. We leave the issues related to building a full model to future works.

Let us consider an antisymmetric $B_{MN}$ tensor field in the gravity background. We introduce the fermion interaction,
\be
 \label{antieqn}
S_1 =   -g_1 \int d^4 x \,\sqrt{-g}     B_{NM} \overline{\spinor}_2  \Gamma^{NM} \spinor_1 +
h.c.
\ee
For this term to be gauge invariant, the antisymmetric field must have twice the charge of the spinor, {\it i.e.} $2q$.

In the following, we will assume that the spatial tensor components condense.
The qualitative features of the results will not depend on the exact details of the $B_{xy}(r)$ profile. We will not consider the back-reaction of $B_{xy}$ on the metric.

In order to obtain (approximate) analytical results, we consider a simplified system with $g^{xx}=g^{yy}\equiv h(r)$. We let $W_y$ condense. For simplicity, consider spinor momentum near a Fermi surface, along the $x$ direction: $k_x = k_F + k_\perp$. We turn on a finite coupling to the condensed $B_{xy}$ field, but treat $\eta$, $k_\perp$ and $\omega$ as small perturbations in the Dirac equation. Both the equation of motion and the boundary condition for a charged bosonic probe are real at zero frequency. Thus, for simplicity, we will consider real $g_1$, $B_{xy}$ in the following.

Let us collect the coupled two-component spinors of opposite charges into the {\it modified} Nambu-Gor'kov spinor $\Psi \equiv \binom{\lambda_1}{\sigma^3 \chi_2}$. The Pauli matrix $\sigma^3$ is included in order to make the equations more symmetric: it changes $D_r(-k, -q)$ into $D_r(k, q)$. The indices $1,2$ on $\lambda$ and $\chi$ will be suppressed in the following. In order to simplify the computation, we use a basis of ingoing solutions ($\bI, \bII$) having $\chi^\bI  = 0$ and $\lambda^\bII  = 0$, respectively, when the $\eta$ and $g_1$ couplings are turned off.

\subsection{Finite B-field background}

If a finite, possibly large, B-field is turned on, the ingoing wavefunctions change in the following way,
{\footnotesize
\be \nonumber
   \Psi^{\bI} =  \binom{ \lambda^{\bI}}{0} \rightarrow  \tilde\Psi^{\bI}  =
   \binom{\tilde \lambda^{\bI}}{\sigma^3 \tilde\chi^{\bI}},
\quad
    \Psi^{\bII} =  \binom{0}{\sigma^3 \chi^{\bII}} \rightarrow  \tilde\Psi^{\bII}  =
   \binom{\tilde \lambda^{\bII}}{\sigma^3 \tilde\chi^{\bII}} .
\ee }

\vskip 0.4cm

\noindent
The Dirac equation for the modified Nambu spinor,
\begin{equation}
\label{ceqn1}
\begin{pmatrix} D_r (k, q) &  - i g_1 h(r) B_{xy}(r) \sigma^3   \\
i g_1 h(r) B_{xy}(r) \sigma^3  &  D_r(k, q) \end{pmatrix}
 \tilde \Psi^{\bI}(r)   =0
\end{equation}
and similarly for $\tilde \Psi^{\bII}$.
Tilde will indicate $g_1 \ne 0$ solutions.
If the $g_1=0$ solutions $\lambda^{\mathcal{A}}, \chi^{\mathcal{A}} $ (satisfying $\lambda^{\bI} = \sigma^3 \chi^{\bII} $) are known, then (\ref{ceqn1}) can surprisingly be solved by setting
{\footnotesize
\be
  % \nonumber
  \label{roteq}
  \tilde \Psi^{\mathcal{A}} = \binom{\tilde \lambda^{\mathcal{A}}}{\sigma^3 \tilde\chi^{\mathcal{A}}}  =
   \underbrace{
   \begin{pmatrix}
   \cosh \, b(r) & \ -i \sinh \, b(r)      \\
   i \sinh \, b(r)  &  \ \cosh \, b(r)
   \end{pmatrix}
   }_{X(r)}
   \cdot   \binom{ \lambda^{\mathcal{A}}}{\sigma^3\chi^{\mathcal{A}}}
\ee
}
where $ b(r) =   g_1 \int^r_{r_*} dr' \sqrt{g_{rr}} h(r') B_{xy}(r') $ and $\mathcal{A}={\bI},{\bII}$.  We will use the notation, $X_\infty \equiv X(r\rightarrow \infty)$ and $b_\infty \equiv b(r\rightarrow\infty)$. At finite temperature, the limit is convergent because $B_{MN}$ has no boundary sources and the horizon provides an IR cutoff. (At zero temperature, $b_\infty$ may diverge.)
% which renders the function normalizable.
Compared to the $g_1=0$ case, the $2\times 2$ expectation value matrix in  (\ref{eq:amatrix}) gets multiplied by $X_\infty$ from the left. 

%thus it is a normalizable function.

\subsection{Perturbation theory}

After turning on a finite $B_{xy}$, let us perturb the system by a small $\omega, \eta$ and $k_\perp$. The wavefunctions change,
\be
   \nonumber
   \tilde\Psi^{\mathcal{A}}  \rightarrow  \tilde\Psi^{\mathcal{A}}+\delta \tilde\Psi^{\mathcal{A}} =
   \binom{\tilde \lambda^{\mathcal{A}}}{\sigma^3\tilde\chi^\mathcal{A}} +   \binom{\delta\tilde\lambda^{\mathcal{A}}}{\sigma^3\delta \tilde \chi^{\mathcal{A}}} .
\ee
Here  $\delta \tilde\Psi^\mathcal{A}$ is a small perturbation to the rotated $\tilde \Psi$
wavefunction. Plug this ansatz into the Dirac equation to get,
{\footnotesize
\bea
\label{perteq}
% \nonumber
 &
 \underbrace{ \begin{pmatrix} D_r (k, q) &   - i g_1 h B_{xy} \sigma^3 &  \\
 i g_1 h B_{xy} \sigma^3  &  D_r(k, q) \end{pmatrix} }_{\mathcal{D}}
 \delta  \tilde\Psi^{\mathcal{A}}  \qquad \qquad& \\
& +   \begin{pmatrix} \sqrt{g^{tt}} \omega \sigma^1 -\sqrt{h}k_\perp i\sigma^2  &
 \ -\eta \sqrt{h} W_y \sigma^1   \\
 \eta \sqrt{h} W_y  \sigma^1     &
 \  -\sqrt{g^{tt}} \omega \sigma^1 -\sqrt{h}k_\perp i\sigma^2
 \end{pmatrix}  \tilde \Psi^{\mathcal{A}} =0  &
\nonumber
\eea
}
Let us now integrate the equation using $\int_{r_*}^\infty dr \sqrt{g_{rr}}
\, \overline{\tilde\Psi^{\mathcal{B}} }$. After integration by parts, the differential operator $\mathcal{D}$ will act on the integrand $ \overline{\tilde\Psi^{\mathcal{B}}}$ and vanish. Thus, the only contribution from $\mathcal{D}\delta  \tilde\Psi^{\mathcal{A}} $ % the first term in (\ref{perteq})
will come from the Wronskians computed at the boundary ($\delta \tilde\lambda$, $\delta \tilde\chi$ vanish at the horizon),
\be
 \nonumber
  W^{\mathcal{BA}} = \overline{\tilde\Psi^{\mathcal{B}} } \Gamma^{\underline{r}} \, \delta \tilde\Psi^{\mathcal{A}}= -(\tilde\lambda^\mathcal{B})^\dagger \sigma^2 \delta \tilde\lambda^\mathcal{A} + (\tilde\chi^\mathcal{B})^\dagger \sigma^2 \delta \tilde\chi^\mathcal{A}.
\ee

\subsection{Zero temperature}

At zero temperature at the Fermi surface, the source component of the spinor vanishes ($\tilde\lambda^\mathcal{A}_+=\tilde\chi^\mathcal{A}_+=0$). Thus, %, and thus
%the Wronskians give
\be
 \nonumber
  W^{\mathcal{BA}} =  i(\tilde\lambda^\mathcal{B}_-)^* \, \delta \tilde\lambda^\mathcal{A}_++  i(\tilde\chi^\mathcal{B}_-)^* \, \left(-\delta \tilde\chi^\mathcal{A}_+\right).
\ee
%$ X_\infty^T =  (X_\infty)^*$ and
For simplicity, we will pretend that $|b_\infty|<\infty$ at $T=0$.
Using (\ref{roteq}) and thus $\overline{\tilde\Psi^{\mathcal{B}} } \Gamma^{\underline{r}} \, \delta \tilde\Psi^{\mathcal{A}}= \overline{ \Psi^{\mathcal{B}} } \Gamma^{\underline{r}} \, X_\infty \delta \tilde\Psi^{\mathcal{A}}$, this can

\bwt

\begin{figure}[h!]
\begin{center}
\includegraphics[scale=0.68]{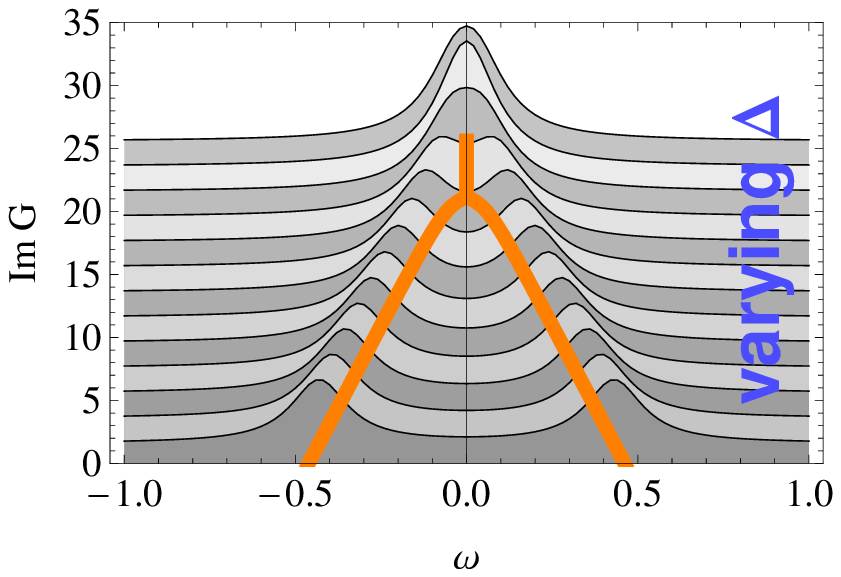}
\includegraphics[scale=0.68]{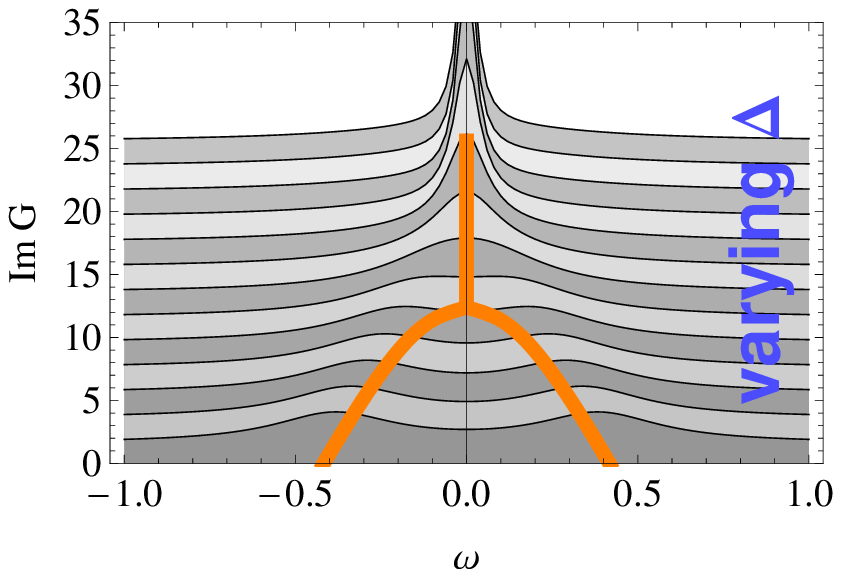}
\includegraphics[scale=0.68]{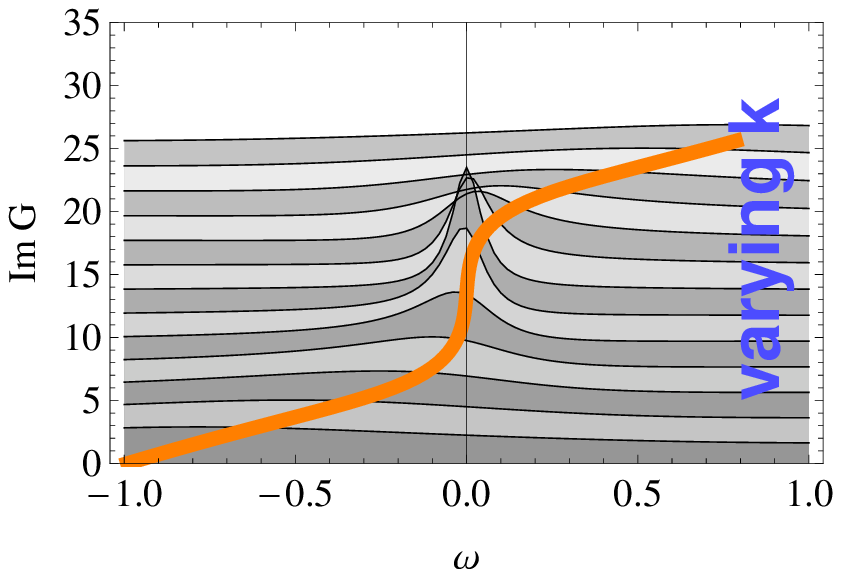}
\caption{Energy distribution curves (EDC) from the formula $G \sim \frac{v_F k_\perp+\omega +i \Gamma \cosh(2 b_\infty)}{v_F^2 k_\perp^2+\Gamma ^2+\Delta ^2-\omega^2-2 i \Gamma  \omega  \cosh(2
b_\infty)}$. \, {\it Left plot:} The various gray curves are EDCs with varying gap parameter $\Delta = 0\ldots 0.5$. The coupling $b_\infty$ is set to zero, and the Green's function simplifies to the BCS case. Orange lines indicate the location of the peak maximum. At small values, temperature broadening ($\Gamma = 0.1$) eliminates the gap. \, {\it Middle plot:} Same EDCs at finite value of antisymmetric coupling ($b_\infty = 0.7$). The gap now vanishes below a larger cutoff. Compared to BCS, the size of the peak is smaller (larger) in the gapped (arc) region. \, {\it Right plot:} The curves here have varying momenta $k_\perp = -1\ldots 1$, with fixed $\Delta = 0.3$ and $\Gamma=0.05$. The antisymmetric coupling is relatively large, $b_\infty = 1.8$. The gap
vanishes and the dispersion is non-linear.
}
\label{panelfig}
\end{center}
\end{figure}

\ewt

 \noindent be further written in matrix form as,
\be
 \nonumber
  W  =  i(\lambda^\bI_-)^* \, X_\infty  \cdot
  \underbrace{
  \begin{pmatrix} \delta \tilde\lambda^\bI_+  &  \delta \tilde\lambda^\bII_+ \\
 -\delta \tilde\chi^\bI_+  &  -\delta \tilde\chi^\bII_+ \end{pmatrix}}_{ \sigma^3 \mathcal{S}_{0}}.
\ee

\noindent
Here we have defined $\mathcal{S}_0$ to be the matrix without the minus signs. The null subscript refers to zero temperature.
Since at $T=0$ we have  $\tilde\lambda^\mathcal{A}_+=\tilde\chi^\mathcal{A}_+=0$,  the source matrix in  (\ref{eq:amatrix}) is in fact equal to $\mathcal{S}_0$.
Integration of the perturbed Dirac equation (\ref{perteq}) gives %The integrated equation of motion yields
\be
  X_\infty \sigma^3 \mathcal{S}_0 %G_{0}^{-1} B_0
  = c_0 % \ \sim \
  X_\infty %\cdot
  \underbrace{
  \begin{pmatrix}
  {\omega }  + v_F k_\perp  & \Delta   \\
  -\Delta   & -{\omega }  + v_F k_\perp
  \end{pmatrix}}_M
  X_\infty
   \nonumber
\ee
The RHS comes from the second term in (\ref{perteq}).
The gap parameter $\Delta$ is proportional to $\eta$, $v_F$ is the Fermi velocity and $c_0$ is a constant. Multiplying from the left by $X_\infty^{-1} = X_\infty^T$ we get,
\be
  \sigma^3 \mathcal{S}_0  = c_0  M X_\infty .
   \nonumber
\ee
The source matrix is related to the Green's function through $G_{0}^{-1} = \mathcal{S}_0 \mathcal{E}_0^{-1}$ where $\mathcal{E}_0$ is the expectation value matrix. The quasiparticle poles are located where $\det \mathcal{S}_0 =0$. Since $\mathcal{S}_0$ is only rotated compared to the $g_1=0$ case, the poles  remain at the same place in the complex $\omega$ plane.
%One obtains the BCS dispersion,
%$  \omega_* = \pm \sqrt{\Delta^2 + v_F^2 k_\perp^2 }$.
%  \nonumber
%\ee

\subsection{Finite temperature}

At small temperatures, the pole in the Green's function at the Fermi surface is not at $\omega=0$, but it is located on the lower half plane. In the spectral function this manifests itself as temperature broadening. The original ($k_\perp=\omega=\eta=0$ and $g_1 = 0$) spinor wavefunctions $\lambda, \chi$ now have a non-zero source component at the $AdS_4$ boundary. Let us denote it by  $i\gamma := \lambda^\bI_+ = \chi^\bII_+ $. We assume that $T \ll \mu$ and thus $|\gamma| \ll 1$.
After integrating the (\ref{perteq}) Dirac equation, the Wronskians give
\bea
 \nonumber
  & W^{\mathcal{BA}} =  i(\tilde\lambda^\mathcal{B}_-)^* \, \delta \tilde\lambda^\mathcal{A}_+ +  i(\tilde\chi^\mathcal{B}_-)^* \, \left(-\delta \tilde\chi^\mathcal{A}_+\right)
 \qquad & \\
 & \qquad \qquad - i(\tilde\lambda^\mathcal{B}_+)^* \, \delta \tilde\lambda^\mathcal{A}_- -  i(\tilde\chi^\mathcal{B}_+)^* \, \left(-\delta \tilde\chi^\mathcal{A}_-\right) .
 &
 \nonumber
\eea
Since $\tilde\lambda^\mathcal{B}_+$ and $\tilde\chi^\mathcal{B}_+$ are proportional to $\gamma$, they are much smaller than $\tilde\lambda^\mathcal{B}_-$ and $\tilde\chi^\mathcal{B}_-$. Thus, we can neglect the second line. Similarly to the $T=0$ case, the Wronskians give,
\be
  W \approx  i(\lambda^\bI_-)^* \, X_\infty \sigma^3 \mathcal{S}_0  .
\ee
To first order, the integrated equation (\ref{perteq}) still gives, %we still have
\be
   \sigma^3 \mathcal{S}_0  \approx c_0 % \ \sim \
   M %\cdot
  X_\infty .
  \label{a0expr}
%   \nonumber
\ee

\noindent
At finite temperatures, however, the source matrix in     (\ref{eq:amatrix})    will be different from $  \mathcal{S}_0 $. Let us denote it by $\mathcal{S} $,
\be
 \label{sdefeq}
 \mathcal{S} = \underbrace{
 \begin{pmatrix} \tilde\lambda^\bI_+  &
 \tilde\lambda^\bII_+  \\
 \tilde\chi^\bI_+    &  \tilde\chi^\bII_+  \end{pmatrix}
 }_{\mathcal{S}'}
 +
  \begin{pmatrix}  \delta \tilde\lambda^\bI_+  &
    \delta \tilde\lambda^\bII_+ \\
   \delta \tilde\chi^\bI_+  &    \delta \tilde\chi^\bII_+ \end{pmatrix} .
\ee
The second matrix is just $\mathcal{S}_0$.
Using (\ref{roteq}), the first matrix can be written as,
\be
 \nonumber
  \sigma^3 \mathcal{S}' =  X_\infty
   \begin{pmatrix} \lambda^\bI_+  & 0 \\
   0   &  -\chi^\bII_+  \end{pmatrix}
 =  i\gamma X_\infty    \sigma^3
 =   i\gamma \sigma^3 X_\infty^T  .
\ee
Thus, (\ref{a0expr}) and (\ref{sdefeq}) combines into,
\be
  \mathcal{S}  \approx \left(c_0  \sigma^3 M \right) X_\infty + (i\gamma) X_\infty^T .
%  \mathcal{S}  \approx \left[c_0  \sigma^3 M \right] X_\infty + [i\gamma] X_\infty^T .
   \nonumber
\ee
Importantly, the perturbations, which are encoded in $M$, are rotated in the opposite direction compared to the width term.  This implies that the antisymmetric coupling does alter the finite temperature correlators.

\begin{figure}[h!]
\begin{center}
\includegraphics[scale=0.8]{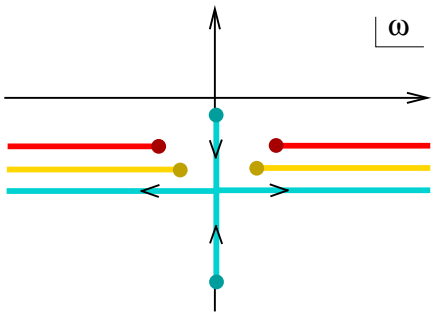}
\caption{Motion of poles of $G(\omega)$ from eqn. (\ref{greenfunceqn}) when momentum is varied. The values of $\Gamma$, $\Delta$ and $v_F$ are fixed. \, (i) Red lines indicate the two poles in the BCS case: they approach the imaginary axis and turn back at a finite distance when $k_\perp$ changes sign (dots indicate this point). \, (ii) Yellow lines show the paths when the fermions are coupled to the antisymmetric field. The gap is reduced because the poles move closer to the imaginary axis. Note that these quasiparticles have larger widths compared to BCS. \, (iii) Blue lines correspond to higher values of the $b_\infty$ coupling. The poles collide at $k_\perp = \pm {1\over v_F} \sqrt{ \Gamma^2 \sinh^2(2 b_\infty)-\Delta^2 } $. For smaller values of $|k_\perp|$, they remain on the negative imaginary axis. Arrows show their motion for $k_\perp>0$.} \label{polemotionfig}
\label{polefig}
\end{center}
\end{figure}

The Green's function can be computed from $G = \mathcal{E}\mathcal{S}^{-1}$ where $\mathcal{E}$ is the diagonal expectation value matrix of the $g_1 = 0$ solution, rotated by $X_\infty$,
\be
  \mathcal{E}  = X_\infty
  \begin{pmatrix} \lambda^{\bI}_- &   0 \\
   0 & \chi^{\bII}_-  \end{pmatrix} = \lambda^{\bI}_- X_\infty
\ee

The Green's function matrix,
{\footnotesize
\bea
 \nonumber
 &
 G \sim  \begin{pmatrix}
    {- v_F k_\perp+\omega +i \Gamma
\cosh 2 b_\infty} &    {-\Delta + \Gamma \sinh 2 b_\infty
}  \\
    {-\Delta - \Gamma \sinh 2 b_\infty
}  &  { v_F k_\perp+\omega +i \Gamma
\cosh 2 b_\infty}  \end{pmatrix} \times
& \\
 & \times\left[v_F^2 k_\perp^2+\Gamma ^2+\Delta ^2-\omega^2-2 i \Gamma  \omega  \cosh 2 b_\infty \right]^{-1} &
  \label{greenfunceqn}
\eea
}
where $\Gamma(T) = \gamma / c_0 \sim T$. The function $b_\infty = b_\infty(T)$ may diverge as $T\rightarrow 0$.
The location of the two poles is readily computed,
\be
  \omega_* = -i \Gamma \cosh(2 b_\infty) \pm \sqrt{\Delta^2 + v_F^2 k_\perp^2 - \Gamma^2 \sinh^2(2 b_\infty)} .
   \nonumber
\ee
The Green's function and the dispersion relation are our central technical results. In the following, we will analyze their properties and show how they can give Fermi arcs.

\subsection{Properties of the Green's function }

\ben

\item

At $b_\infty = 0$, the familiar BCS formula is recovered,
\be
 \nonumber
  G^{-1} \sim {v_F k_\perp-\omega -i \Gamma +\frac{\Delta^2}{\omega +v_F k_\perp+i \Gamma }} .
\ee

\item
The $T \rightarrow 0$ limit also gives back the BCS Green's function unless $b_\infty(T)$ diverges too quickly.

\item

When $b_\infty$ is large (with fixed $\Gamma$), the peak is stuck near $\omega=0$ and disperses very slowly. On the complex $\omega$ plane, the quasiparticle poles move on the negative imaginary axis.

\item

By extracting the imaginary part from (\ref{greenfunceqn}), one can show that the spectral function is always positive. The poles always stay on the lower half plane.

\item

FIG. \ref{panelfig} shows plots of Im $G_{22}(\omega)$.
The left plot shows the BCS case where the various curves have different $\Delta = 0\ldots 0.5$.
At small $\Delta$, temperature broadening kicks in and the maxima of the two peaks coalesce as shown by the orange lines.

The middle plot shows the same figure at finite value of $b_\infty$. The peaks at larger $\Delta$ become wider and at small $\Delta$ taller. The gap now vanishes even at intermediate values of $\Delta$.

In case of a p-wave gap, $\Delta \sim \cos(\theta)$ and thus different values of the $\Delta$ parameter correspond to different angles in momentum space. The second figure thus shows that there is an extended ``Fermi arc'' region where the gap vanishes. This arc is longer than what is justified by temperature broadening.

Finally, the third figure shows the dispersion of the peak at large $b_\infty$. The gap
vanishes and the dispersion is non-linear,  ${d\omega_\star \over dk} < v_F$.

\item

FIG. \ref{polefig} shows the paths of the poles in the Green's function as the momentum is varied. This figure has been separately confirmed by numerical computations (using a real B-field profile). For smaller values of $b_\infty$, the size of the gap decreases.
The effective quasiparticle width is $\Gamma_{eff} = \Gamma \cosh(2 b_\infty) > \Gamma$.
For large enough $b_\infty$, the poles actually collide and then move on the imaginary axis.

\item

When $\omega=0$, the spectral function simplifies,
\be
\nonumber
  \Im G \sim \frac{  \Gamma
\cosh 2 b_\infty}{v_F^2 k_\perp^2+\Gamma ^2+\Delta ^2} .
\ee
Since $b_\infty$ only appears through a multiplicative factor, it does not have a significant effect on an $\omega=0$ two-dimensional ARPES-type figure of the ``Brillouin zone''. Any visible arcs in such a figure will be similar to arcs caused by temperature broadening (see FIG. \ref{tripfig} (i)).

\item

\begin{figure}[h!]
\begin{center}
\includegraphics[scale=0.70]{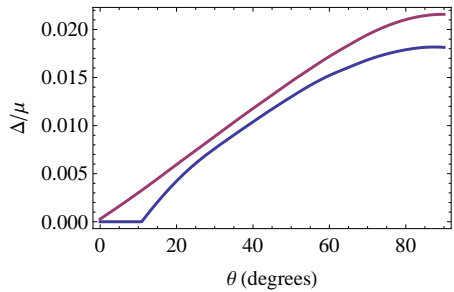}
\caption{Sample numerical plot for the gap as a function of the angle. The gap is defined as half the minimal distance of peak maxima in the spectral function. (i) Purple curve shows the p-wave gap. The effects of temperature broadening are invisibly small at this temperature. (ii) Blue curve corresponds to a finite value of the antisymmetric coupling.
The parameters of the background: $\mu =5.12$, $q=1$, $m_W^2 = -0.2$, $|k_F(\theta)| = 2.20$ (at $0^\circ$)\ldots 2.27 (at $90^\circ$). } \label{gapfig}
\end{center}
\end{figure}

Even though the (normalized) spectral function does not change at $\omega=0$, the gap does vanish in an extended arc region.
FIG. \ref{gapfig} shows the gap as a function of the angle, computed numerically. Purple curve shows the p-wave gap. The temperature is rather small and its broadening effects cannot be seen. There are, however, larger, visible effects when the B-field is turned on (blue curve): the gap vanishes for $|\theta| \lesssim 10^\circ $ in the Fermi arc region.

\een

%\clearpage

\bwt

\begin{figure}[h!]
\begin{center}
\includegraphics[scale=0.85]{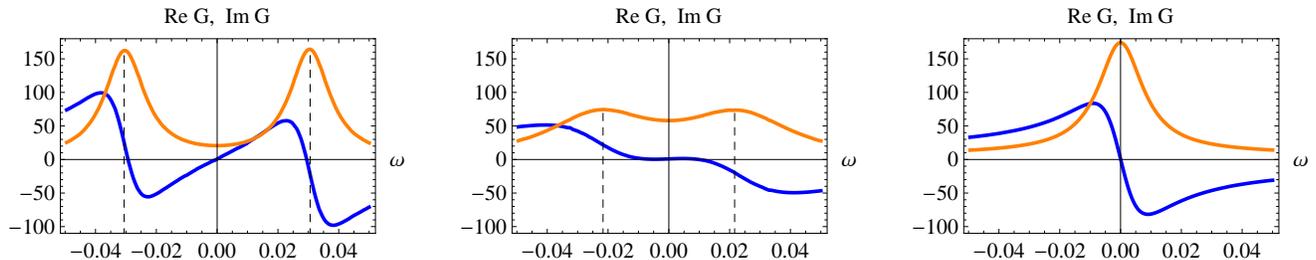}
\caption{Sample Green's functions computed numerically on a p-wave background (parameters: $\mu =5.12$, $q=1$, $m_W^2 = -0.2$). The real and imaginary parts are plotted with blue and orange colors, respectively.  {\it Left plot:} EDC generated at $|k| = 2.208$ at a $20^\circ$ angle away from the gapless point. $|k|$ has been chosen such that the peaks are closest to each other. Thus, their distance equals to the gap, $\Delta \sim 0.03$.    {\it Middle plot:} A finite antisymmetric coupling reduces the gap,  $\Delta \sim 0.02$. The peaks have a larger width. {\it Right plot:} If the antisymmetric coupling is further increased, the gap vanishes and only one large peak is visible. } \label{greenfig}
\end{center}
\end{figure}

\ewt

\quad
\vskip -1.5cm
\quad
\section{Discussion}

In this paper, we found Fermi arcs in holographic superconductors using a phenomenological approach. A condensing vector field (which could be substituted with higher spin fields) produces an anisotropic gap in the probe fermion spectral functions. A coupling between the fermions and a charged antisymmetric tensor field reduces the size of the gap. If the original gap at a certain point in momentum space was small enough, then it gets completely eliminated and a non-linear dispersion is produced. This happens near the nodal points where the p-wave gap is small. Thus, the gapless points will become a finite length Fermi surface: a Fermi arc.

An approximate analytical formula (\ref{greenfunceqn}) for the fermionic Green's function has been derived and separately confirmed by numerical calculations.
We are currently lacking a purely boundary field theory (or ``semi-holographic'' \cite{Faulkner:2010tq}) interpretation of this phenomenon. % which would be of great value.

Increasing the temperature increases the width of the quasiparticles, but presumably reduces $|b_\infty|$. It would be interesting to construct a full model where the backreaction of the antisymmetric field is also taken into account.
This would allow for a study of the arc length as a function of temperature and other parameters.

In an $SU(2)$ holographic superconductor with spinor doublet $\spinor$, a similar effect may arise from the coupling
\be
  \mathcal{L}_{int} = \overline{\spinor}\Gamma^{MN} F^i_{MN} {\sigma^i \over 2} \spinor . \nonumber
\ee
where $F^i_{MN}$ is the $SU(2)$ field strength.
When expanded, this contains a term
\be
  \mathcal{L}_{int} \supset   \overline{\spinor}_2  \Gamma^{NM} \spinor_1 \left[ \partial_M W^+_N - \partial_N W^+_M \right]   + h.c. \nonumber
\ee
with $W^{\pm}_M = A^1_M \pm i A^2_M$. If there are spatial (amplitude) fluctuations in the order parameter, this term may produce similar effects to the antisymmetric coupling that we introduced in section \ref{sec:bfield}.

In the Appendix, we describe another model which deals with ``spin order'' represented by a neutral scalar field in the bulk. The coupling of this field to the fermions creates a closed Fermi pocket. It would be interesting to build a more realistic holographic model using this idea.

Finally, it is worth emphasizing that in both cases, the system is in the superconducting phase. In real materials, arcs appear in the non-superconducting metallic phase. It would be interesting to understand whether long-range order could be eliminated in holographic systems while preserving Fermi arcs.

\vspace{0.2in}   \centerline{\bf{Acknowledgments}} \vspace{0.2in} I thank Hong Liu for collaboration, many suggestions and comments on the manuscript. I further thank Tom Faulkner and John McGreevy for many comments, and Michal Heller, Nabil Iqbal and M\'ark Mezei for collaborations on related projects.

%\clearpage

\section*{Appendix: Fermi pocket}

In this appendix, we demonstrate how a neutral field corresponding to spin (density wave) order can modify the anisotropic Fermi surface of the fermions.

Let us introduce ``spin'' in the boundary theory by considering two identical bulk Dirac fields, $\psi_\uparrow$ and $\psi_\downarrow$.  Here up and down refer to a new $z$ direction which is perpendicular to the 2d superconductor and is not be confused with the radial direction of $AdS_4$.

We can arrange the spinors into a {\it spin} $SU(2)$ doublet $\psi=\binom{\psi_\uparrow}{\psi_\downarrow}$. Naturally, we only want to consider Lagrangians which are invariant under the spin $SU(2)$.

\begin{figure}[h!]
\begin{center}
\hskip -0.6cm \includegraphics[scale=0.9]{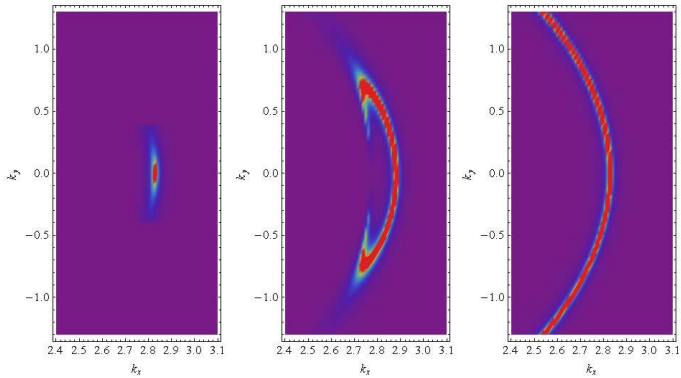} \caption{(i) Fermion response function showing a p-wave gap. The parameters of the background:  $\mu = 6.95$, $q = 1$, $m_W^2 = -0.2$. (ii) Coupling to the neutral field shifts the Fermi momenta of the spin up and spin down fermions in opposite directions. Since the eigenvalue repulsion occurs away from the origin, one of the poles will cross $\omega=0$ thereby creating a Fermi arc. (iii) Fermi surface without any coupling to the condensed fields. } \label{tripfig}
\end{center}
\end{figure}

In order to write down physically interesting interaction terms, it is useful to consider another $SU(2)$ symmetry called the {\it pseudospin}.  The $U(1)$ bulk gauge symmetry may be interpreted as the unbroken diagonal subgroup and the $W$ field is the pseudospin W-boson. The pseudospin $SU(2)$ acts on the doublet
$
   \spinor = \binom{\spinor_1}{ \spinor_2} = \binom{
  \psi_\uparrow}{ (\psi_\downarrow)^c }
$
where $\chi^c = C \Gamma^t \chi^*$ is the charge conjugate spinor. Thus, $\spinor_1$ and
$\spinor_2$ have opposite charges, $\pm q$, under the $U(1)$ gauge symmetry. These are the spinors that we used in the rest of the paper.

The spin order parameter $\Phi^i$ is a spin triplet \cite{Iqbal:2010eh, sachdev}. Its condensate breaks the spin $SU(2)$ down to $U(1)$. Since the fermions form a spin doublet $\psi = \binom{ \psi_\uparrow}{\psi_\downarrow } = \binom{ \spinor_1}{ (\spinor_2)^c }$, a natural coupling between the fields is
\be
   L= i g_\Phi \Phi^i  \overline{\psi}_a  (\sigma^i)_{ab} {\psi}_b .
    \nonumber
\ee

We assume that only $\Phi \equiv \Phi^z$ condenses. Hence, we can rewrite the interaction as $L =  i g_\Phi \Phi \overline{\spinor}_a (\sigma^z)_{ab} {\spinor}_b =  i g_\Phi \Phi \left( \overline{\spinor}_1 {\spinor}_1- \overline{\spinor}_2 {\spinor}_2 \right)$. Here we used the fact that $\bar\chi \chi = \overline{\chi^c} \chi^c$ for a Dirac spinor.

The action then takes the form,
%\vskip -0.5cm
\be
S_1 = \int d^4 x \,\sqrt{-g}  \left[  - \partial_M \Phi \partial^M \Phi - m_{\Phi}^2 \Phi^2 +  i
g_\Phi \Phi  \overline{\spinor}_a  (\sigma^z)_{ab} {\spinor}_b \right]
\nonumber
\ee
\vskip 0.3cm

Note that the two spinors get contributions to their effective masses with opposite signs. As a result, the degeneracy of the Fermi momenta of the two spinors is resolved: $k_F^{(1)} \ne k_F^{(2)}$. Originally, the poles of $G_{\spinor_1}(\omega)$ and $G_{\spinor_2}(\omega)$ collided at $\omega=0$ as the momentum increased. Since the Fermi momenta are different now, one of the peaks will cross $\omega=0$ if the coupling to $\Phi$ is large enough compared to the gap. This results in a Fermi arc.

We emphasize that the arc here is only present if one plots the spin components separately. Since the other spin component has its arc on the other side of the ``banana'', the trace will be a full oval, a Fermi pocket.

The shift in the Fermi momenta can be modeled by the following Green's function,
\be
  G^{-1} = \omega - v_F k_\perp -\varepsilon_1 + i \Gamma - {\Delta^2 \over \omega + v_F k_\perp - \varepsilon_2 + i \Gamma}
  \nonumber
\ee
where $\varepsilon_i / v_F$ are the shifts in the Fermi momenta for the two spinors. Depending on the parameters, this function gives an arc similar to the one in FIG. \ref{tripfig} (ii).

\bibliography{fermiarc}

\end{document}